\documentclass[a4paper,12pt,english]{article}
\usepackage{amsfonts,bm,amssymb,array,babel,cite}
\usepackage[mathscr]{euscript}
\usepackage[all,color]{xy}
\usepackage{relsize}
\usepackage{tikz}
\usepackage{url}
\usepackage{amsmath,amsthm,bm}
\usepackage{cancel}
\usepackage{accents}
\usepackage[]{graphicx}
\usepackage{mwe} % new package from Martin scharrer
\usepackage{caption}
\usepackage[T1]{fontenc}
\usepackage[toc,page]{appendix}

\usepackage{empheq}
\usepackage{color}
\usepackage{mathtools}
\makeatletter

\newcommand{\bbm}{\left(\begin{matrix}}
	\newcommand{\ebm}{\end{matrix}\right)}
\newcommand{\beq}{\begin{eqnarray}}
\newcommand{\eeq}{\end{eqnarray}}

\makeatother

 \def\one{\mbox{1 \kern-.59em {\rm l}}}
\begin{document}
\begin{flushright}
LAPTH-003/19
\end{flushright}
\begingroup
{\let\newpage\relax% Void the actions of \newpage
\title{Four-dimensional Gravity on a Covariant Noncommutative Space}

\author{G. Manolakos\textsuperscript{1},\,P. Manousselis\textsuperscript{1} \,G. Zoupanos\textsuperscript{1,2,3,4}}\date{}
\maketitle}
\begin{center}
\emph{E-mails: gmanol@central.ntua.gr\,, pman@central.ntua.gr\,, George.Zoupanos@cern.ch }
\end{center}

\begin{center}
\itshape\textsuperscript{1}Physics Department, National Technical
University, GR-15780 Athens, Greece\\
\itshape\textsuperscript{2} Institute of Theoretical Physics, D-69120 Heidelberg, Germany\\
\itshape\textsuperscript{3} Max-Planck Institut f\"ur Physik, Fohringer Ring 6, D-80805 Munchen, Germany\\
\itshape\textsuperscript{4}Laboratoire d' Annecy de Physique Theorique, Annecy, France
\end{center}
\vspace{0.1cm} \emph{Keywords}: gauge theory, four-dimensional gravity, covariant noncommutative spaces, fuzzy de Sitter, covariant transformation of field strength

\begin{abstract}
\noindent We formulate a model of noncommutative four-dimensional gravity on a covariant fuzzy space based on SO(1,4), that is the fuzzy version of the $\text{dS}_4$. The latter requires the employment of a wider symmetry group, the SO(1,5), for reasons of covariance. Addressing along the lines of formulating four-dimensional gravity as a gauge theory of the Poincar\'e group, spontaneously broken to the Lorentz, we attempt to construct a four-dimensional gravitational model on the fuzzy de Sitter spacetime. In turn, first we consider the SO(1,4) subgroup of the SO(1,5) algebra, in which we were led to, as we want to gauge the isometry part of the full symmetry. Then, the construction of a gauge theory on such a noncommutative space directs us to use an extension of the gauge group, the SO(1,5)$\times$U(1), and fix its representation. Moreover, a 2-form dynamic gauge field is included in the theory for reasons of covariance of the transformation of the field strength tensor. Finally, the gauge theory is considered to be spontaneously broken to the Lorentz group with an extension of a U(1), i.e. SO(1,3)$\times$U(1). The latter defines the four-dimensional noncommutative gravity action which can lead to equations of motion, whereas the breaking induces the imposition of constraints that will lead to expressions relating the gauge fields. It should be noted that we use the Euclidean signature for the formulation of the above programme.  
\end{abstract}
\vspace{2cm}
\tableofcontents
\vspace{0.5cm}
\hrule

\section{Introduction}
The aim to go beyond the notion of classical spacetime has been with us since 
decades. In particular, there is no obvious physical reason to expect that the 
commutativity of coordinates and the corresponding description of spacetime to 
hold up to arbitrary small length scales. Spacetime noncommutativity, inspired 
by the successes of quantum mechanics and quantum field theory, was suggested 
as their logical extension providing an interesting alternative framework to describe physics close to the Planck scale. As a bonus it was expected that the 
natural appearance of a minimal length scale could act as an effective ultraviolet cutoff in field theories defined on the new spacetime structure \cite{Snyder:1946qz,Yang:1947ud,connes,madorej,Madore:1991bw,buric-grammatikopoulos-madore-zoupanos,filk,grosse-wulkenhaar,grosse-steinacker,connes-lott,martin-bondia,dubois-madore-kerner,DuboisViolette:1988vq,madorejz,connes-douglas-schwarz,seiberg-witten, Banks:1996vh, Ishibashi:1996xs,Aoki:1998vn,Hanada:2005vr,Furuta:2006kk,jurco,chaichian,
camlet,aschieri-madore-manousselis-zoupanos,aschieri-grammatikopoulos,steinacker-zoupanos,chatzistavrakidis-steinacker-zoupanos,fuzzy,Gavriil:2015lka,Szabo:2001kg}. Moreover, on the gravity side, noncommutativity could be used to resolve the singularities of general relativity \cite{Maceda:2003xr, Chamseddine:2016uef}. On the other hand a diffeomorphism-invariant gravity theory is obviously invariant with respect to transformations whose parameters are functions of spacetime, just as in the local gauge theories. Then, naturally, it has been long believed that gravity theory can be formulated as a gauge theory \cite{Utiyama:1956sy,Kibble:1961ba,MacDowell:1977jt,Stelle:1979aj, Ivanov:1980tw,Kibble:1985sn}. In particular in refs \cite{MacDowell:1977jt,Stelle:1979aj} the action obtained is one with cosmological constant, which when positive defines the de Sitter (dS) spacetime to which we refer in the following. Then merging the above two approaches by describing gravity as gauge 
theory on noncommutative spaces is a reasonable consequence. In addition, the prospect of regularizing quantum field theories, or even better, building finite ones are the features that render this approach as a promising 
framework. 

The construction of quantum field theories on noncommutative spaces is a difficult task though and, furthermore, problematic ultraviolet features have 
emerged \cite{filk, grosse-wulkenhaar, grosse-steinacker}. Still noncommutative geometry has been 
proposed as an appropriate framework to accommodate particle models with noncommutative gauge theories \cite{connes-lott} (see also \cite{jurco,chaichian,camlet, Brandt:2003fx,Carlson:2001sw}).
It is worth-noting that a very interesting development in the framework of 
noncommutative geometry is the programme in which extra dimensions of 
higher-dimensional theories are considered to be noncommutative (fuzzy) \cite{aschieri-madore-manousselis-zoupanos,aschieri-grammatikopoulos,steinacker-zoupanos,chatzistavrakidis-steinacker-zoupanos,fuzzy,Gavriil:2015lka}. This programme overcomes the ultraviolet/infrared problematic
behaviours of theories defined on noncommutative spaces. A very welcome feature 
of such theories is that they are renormalizable, contrary to all known
higher-dimensional theories. Even more appealing is that this programme results 
to phenomenologically promising four-dimensional unified theories.

 We recall that a fuzzy two-sphere \cite{Madore:1991bw} (see also \cite{Dolan:2001gn, OConnor:2006iny}) is constructed from finite-dimensional 
matrices and the size of matrices represents the number of quanta on the
noncommutative manifold. The fuzzy sphere, $\text{S}^2_F$, at fuzziness level N-1, is the 
non-commutative manifold whose coordinate functions are N$\times$N matrices 
proportional to the generators of the N-dimensional representation of SU(2). 
Introducing a cutoff parameter N-1 for angular momentum in a two-sphere, the 
number of independent functions is N$^2$. Then one can replace the functions 
defined on this noncommutative
manifold by N$\times$N matrices, and therefore algebras on the sphere become 
noncommutative. A generalization to a higher dimensional sphere is, however, not straightforward. In particular, in the case of a four-dimensional sphere, the same procedure leads to a number of independent functions which is not a square of an integer. Therefore, one cannot construct a map from functions to matrices. One can restate this difficulty algebraically. Algebras of a fuzzy four-sphere have been constructed in \cite{Kimura:2002nq} and the difference from the fuzzy two-sphere case is that the commutators 
of the coordinates do not close in the fuzzy four-sphere
case. This is the source of the difficulties to analyse field theories on the fuzzy four-sphere (see \cite{Kimura:2002nq} and references therein)\footnote{For more details about fuzzy four-sphere see \cite{Medina:2012cs, Medina:2002pc}.}.

 In ref. \cite{Chatzistavrakidis:2018vfi} (see also \cite{Manolakos:2018hvn,Manolakos:2018isw}), we started a programme realizing gravity as a noncommutative gauge 
theory in three dimensions. Specifically, we considered three-dimensional 
noncommutative spaces based on SU(2) and SU(1,1), as foliations of fuzzy 
two-spheres \cite{Grosse:1993uq, Gere:2013uaa} and fuzzy two-hyperboloids \cite{Jurman:2013ota}, respectively. This onion-like 
construction led to a matrix model, which was analysed in a straightforward 
way. In the present work, we would like to follow a different strategy, originally suggested by Yang \cite{Yang:1947ud}, as an extension of the first noncommutative 
model of Snyder \cite{Snyder:1946qz}, which preserves the spacetime isometries\footnote{This idea has been also revived recently by other authors \cite{Heckman:2014xha,Buric:2015wta,Sperling:2017dts} in connection to their research programmes.}. 

Formulating gravity in the noncommutative framework is a rather difficult task, since common noncommutative deformations break Lorentz invariance. However, there are special types of deformations that constitute the \emph{covariant noncommutative spacetimes} \cite{Snyder:1946qz,Yang:1947ud}, which preserve the spacetime isometries. More specifically, Snyder \cite{Snyder:1946qz} was the first to associate the position operators with elements of a Lie algebra in order to construct a Lorentz invariant discrete specetime. Snyder's position operators belong to the four-dimensional de Sitter algebra, SO(1,4). Then, Yang \cite{Yang:1947ud} introduced SO(1,5) in order to allow continuous translations in Snyder's framework and finally Kastrup \cite{Kastrup:1966zzb} considered the four-dimensional conformal algebra SO(2,4). The natural framework of the above early works in the framework of noncommutative geometry, with possible applications in string theory, was explored by Heckman-Verlinde \cite{Heckman:2014xha}, where the authors build a general conformal field theory defined on covariant noncommutative versions of four-dimensional (A)dS spacetime. Therefore, covariant noncommutative spacetimes appear to overcome the problem of breaking Lorentz invariance, which is usually encountered in noncommutative spaces. They preserve Lorentz invariance and, moreover, they provide a short-distance cutoff. In the present work, as a natural further step, we promote the Lorentz invariance of covariant noncommutative spacetime to local symmetry and write down the corresponding gravity theory on such a space. The simplest covariant space is the fuzzy sphere, which however cannot be extended in other cases in a straightforward manner, as we explained above. The suggestion is that in order to formulate a fuzzy covariant noncommutative space, one has to employ a wider symmetry containing all generators of the first one. Then, in practice, in the case of the 
fuzzy de Sitter four-dimensional spacetime, $\text{dS}_4$, that we would like to construct 
here, instead of using its algebra of generators, SO(1,4) we consider a larger 
one in which the latter can be embedded\footnote{The SO(5)$\sim$SO(1,4) in the present formalism, would have the same problem in its description by N$\times$N matrices, as in the fuzzy four-sphere case.}. Such algebras are the SO(1,5) or the conformal, SO(2,4). The larger algebra, say SO(1,5) contains all the generators of the subalgebra SO(1,4) and the commutation relations of the latter are covariant as seen in the larger group.

The next important point is that the coordinates and other operators, too, can be 
represented by N$\times$N matrices in some higher-dimensional representation of the 
enlarged algebra. Of course in this procedure one has to identify the 
various generators of the algebras with operators corresponding to various 
observables such as coordinates, momenta, angular momenta etc. To realize the above picture of a fuzzy covariant noncommutative spacetime in 
the present work we consider the four-dimensional fuzzy de Sitter space of a 
given radius embedded in a five-dimensional spacetime.
However in our construction we describe four-dimensional gravity by
gauging the SO(1,4) subgroup of SO(1,5)\footnote{For an alternative approach (based on the star product) for the gauging of the same group in the noncommutative regime to formulate gravity, see \cite{Chamseddine:2002fd,Chamseddine:2000si}}. To facilitate the algebra we construct the whole model in the Euclidean signature. Therefore, we start with a space with isometries given by the SO(5) group, extend, for reasons of covariance, to the SO(6) symmetry  and then we gauge its maximal SO(5) subgroup to formulate gravity as a gauge theory on the above space. The results we obtain could be reformulated to the Lorentzian signature. It should be noted that we do not approximate manifolds by N$\times$N matrices. We approximate fields defined on manifolds by N$\times$N matrices. In addition, using the tools of noncommutative geometry, we formulate differential geometry on a particular space of N$\times$N matrices, see e.g. \cite{DuboisViolette:1988vq}.

Finally, we should stress that field theories on noncommutative spaces are not quantum theories, see e.g. \cite{Chamseddine:2015ata}, where the quantized version of Connes-Lott approach to standard model is discussed. Similarly, a gravity theory on noncommutative spacetime is not a theory of quantum gravity. We have reasons to expect (in the present Euclidean approach related to the finite degrees of freedom) and hope that the quantized version will be well defined or even finite. Eventually, our aim is to formulate a gravity theory in the general framework of noncommutative geometry but the generalization in this framework of Riemannian geometry is not obvious and is under research by mathematicians \cite{Fathizadeh:2019rzj}. Therefore, in the present paper, we try to give a more down to earth approach via gauge theory, given the fact that gauge theory on noncommutative spaces is well formulated. Therefore, we feel it is natural to use the gauge approach to noncommutative gravity. In our formulation on the noncommutative space, fields are fluctuating around the Euclidean version of Snyder-Yang space as can be read by the covariant coordinate. Using gauge theory we derive the action \eqref{action48}, which can be interpreted as a new and non-trivial matrix model. To formulate quantum theory of gravity on noncommutative spaces is an open and interesting project and probably there are connections to Matrix theory approach, see e.g. \cite{Witten:2001kn,Ydri:2017ncg} where matrix models are studied. Also, see refs. \cite{Steinacker:2019dii,Sperling:2019xar} for the problem of emergent gravity.    

The outline of the paper is as follows. In sections 2, 3 we briefly present the four-dimensional Einstein and conformal gravity theories for reasons of completeness but also, for reference on some of their important features in the next sections. However, we should note that in our present work we do not aim at a review of the framework of noncommutative geometry. For this purpose, we refer to review articles throughout the paper, especially in the beginning of this introduction. In section 4, we present the construction of the four-dimensional fuzzy space on which we formulate gravity as a (noncommutative) gauge theory and describe the construction of the gauge theory of the isometry group of the above space.  In section 5, we comment on the constraints employed in order to break the initial symmetry to the desirable one - the Lorentz. In section 6, we propose an action and comment on the obtaining of the equations of motion. Eventually, in section 7, we write our conclusions and stress out the main features of our model. In Appendix A we give some details on how to deal with the general problem of the (non-) covariance transformation property of the field strength tensor on such covariant spaces and how this can be relieved, while in Appendix B, we give the calculations and results of the transformations of the fields and the component tensors.

\section{Four-dimensional Einstein gravity as a gauge theory}\label{Einstein's gravity}

An alternative way to recover the celebrated results of the theory of general relativity is to approach the whole concept from a gauge-theoretic point of view. This undertaking began in the middle 50's \cite{Utiyama:1956sy} and kept the interest of the physicists for the next three decades \cite{Kibble:1961ba, MacDowell:1977jt, Stelle:1979aj, Ivanov:1980tw, Kibble:1985sn}.
Gauging the Poincar\'e group leads to the Einstein-Cartan theory \cite{Hehl:1976kj}. In order to recover the dynamics of Einstein's gravity, one has to employ the four-dimensional de Sitter group, SO(1,4) and introduce a scalar field in its fundamental representation. The latter, with the appropriate potential, induces a spontaneous symmetry breaking which leads to the Einstein-Hilbert action, starting from an action of Yang-Mills type \cite{Stelle:1979aj,Kibble:1985sn}. In ref. \cite{MacDowell:1977jt}, a similar approach is followed gauging the de Sitter group and considering constraints on the curvature tensors for the final action (see also \cite{Freund:1986ws}). Accordingly, four-dimensional Einstein's gravity can be formulated as a gauge theory, but not as a pure ISO(1,3) gauge theory. Here we recall the whole idea since it is of central importance in our present project. Therefore, in this section, we briefly present the whole construction of the four-dimensional Einstein gravity as a gauge theory, in order to make this project complete and self-contained.    

In the gauge-theory approach of the four-dimensional gravity, at first, one has to consider the vielbein formulation of general relativity. The gauge theory is constructed on the four-dimensional Minkowski spacetime, $\text{M}^4$, and the gauge group is the Poincar\'e group, ISO(1,3). The choice of the Poincar\'e group as the symmetry group is very reasonable, as it is the isometry group of the Minkowski spacetime. The generators of the group satisfy the following commutation relations:  

\begin{equation}
[M_{ab},M_{cd}]=4\eta_{[a[c}M_{d]b]}~,\,\,\,\,\,
[P_a,M_{bc}]=2\eta_{a[b}P_{c]}~,\,\,\,\,\,
[P_a,P_b]=0~,
\end{equation}
where $\eta_{ab}$ is the (mostly positive)  Minkowski metric, $M_{ab}$ are the Lorentz group generators and $P_a$ are the generators of the translations. According to the gauging procedure, one has to introduce the gauge potential, $A_\mu$, which can be decomposed on the generators of the algebra, accompanied by functions that are identified as the gauge fields. Specifically in this case, these functions are the vierbein, $e_\mu^{~a}$ and the spin connection, $\omega_\mu^{~ab}$, which correspond to the translations, $P_a$, and Lorentz generators, $M_{ab}$, respectively. The gauge connection is decomposed as follows:
\begin{equation}
A_{\mu}(x)=e_{\mu}{}^a(x)P_a+\frac{1}{2}\omega_{\mu}{}^{ab}(x)M_{ab}~.\label{connection}
\end{equation}
The above gauge connection is assigned in the adjoint representation, which means that it transforms according to the following rule: 
\begin{equation}
\delta A_{\mu}=\partial_{\mu}\epsilon+[A_{\mu},\epsilon]~,\label{transformconnection}
\end{equation}
where $\epsilon=\epsilon(x)$ is the gauge transformation parameter that is also expanded on the ten generators of the algebra:
\begin{equation}
\epsilon(x)=\xi^a(x) P_a+\frac{1}{2} \lambda^{ab}(x)M_{ab}~.\label{parameter}
\end{equation}
The transformations of the component gauge fields are given after combining the expressions \eqref{connection} and \eqref{parameter} with \eqref{transformconnection}. The resulting expressions for the transformations of the vierbein and the spin connection are: 
\begin{align}
\delta e_{\mu}{}^a&=\partial_{\mu}\xi^a+\omega_{\mu}{}^{ab}\xi_b-\lambda^a{}_be_{\mu}{}^b~,
\\
\delta \omega_{\mu}{}^{ab}&=\partial_{\mu}\lambda^{ab}-2\lambda^{[a}{}_c\omega_{\mu}{}^{cb]}~.
\end{align}
The corresponding field strength tensor (the curvature) of the gauge theory is defined as:
\begin{equation}
R_{\mu\nu}(A)=2\partial_{[\mu}A_{\nu]}+[A_{\mu},A_{\nu}]~.\label{usualformula}
\end{equation}
Since the field strength tensor takes also values in the algebra that is gauged, it can be also expanded on its generators:
\begin{equation}
R_{\mu\nu}(A)=R_{\mu\nu}{}^a(e)P_a+\frac{1}{2} R_{\mu\nu}{}^{ab}(\omega)M_{ab}~,\label{expansion_of_ curvature}
\end{equation}
where $R_{\mu\nu}{}^a(e)$ and $R_{\mu\nu}{}^{ab}(\omega)$ are the component curvatures of the  gauge fields, which are identified as the torsion and curvature, respectively. Their expressions are obtained by replacing equations \eqref{connection} and \eqref{expansion_of_ curvature} into the \eqref{usualformula}: 
\begin{align}
R_{\mu\nu}{}^a(e)&=2\partial_{[\mu}e_{\nu]}{}^a-2\omega_{[\mu}{}^{ab}e_{\nu]b}~,\\
R_{\mu\nu}{}^{ab}(\omega)&=2\partial_{[\mu}\omega_{\nu]}{}^{ab}-2\omega_{[\mu}{}^{ac}\omega_{\nu]c}{}^b~.\label{curvtwoform}
\end{align}
As for the dynamic part of the theory, one would consider the obvious choice of the Yang-Mills action of the Poincar\'e group. However, since the intention is to end up with the action of the Einstein gravitational theory, the Poincar\'e symmetry has to be broken to the Lorentz. Spontaneous symmetry breaking is the indicated way to realize it and can be induced by the inclusion of a scalar field which belongs to the fundamental representation of the SO(1,4) \cite{Stelle:1979aj, Ivanov:1980tw}. The latter (de Sitter group) can be equally chosen instead of the Poincar\'e, and for the present purpose, i.e. the breaking of the initial symmetry to the Lorentz, it is preferred, since it is a semisimple group and all generators can be considered on equal footing, i.e. use a single gauge index to describe all ten generators. The spontaneous symmetry breaking breaks the four generators that are related to the translations, resulting to a constrained theory with vanishing torsion and the SO(1,4) Yang - Mills action considered at first, reduced to an action including the Ricci scalar and preserving only the Lorentz symmetry. This action is indeed the desired Einstein-Hilbert action.

Summing up, Einstein four-dimensional gravity can be considered as a gauge theory of the Poincar\'e algebra up to retrieving the results of the transformations of the fields and the expressions of the curvature tensors. For the dynamic part, in order to obtain the correct action, one has to begin with a Yang - Mills action obeying the de Sitter symmetry, instead of the Poincar\'e, and include a scalar field for inducing a spontaneous symmetry breaking which eventually leads to the Einstein-Hilbert action. 

An equivalent argument leading to the desired, Lorentz symmetry, is to demand the gauge symmetry of the vacuum  to be the SO(1,3), Lorentz group. Therefore, the field strength tensor that corresponds to the generators that break, i.e. the translations, has to be vanishing, recovering in this way the torsionless condition. This condition plays the role of the constraint that is necessary in order to result with the desired symmetry. From this constraint, one obtains an expression of the spin connection as a function of the vielbein: 
\begin{equation}
\omega_\mu^{~ab}=\frac{1}{2}e^{\nu a}(\partial_\mu e_\nu^{~b}-\partial_\nu e_\mu^{~b})-\frac{1}{2}e^{\nu b}(\partial_\mu e_\nu^{~a}-\partial_\nu e_\mu^{~a})-\frac{1}{2}e^{\rho a}e^{\sigma b}(\partial_\rho e_{\sigma c}-\partial_\sigma e_{\rho c})e_\mu^{~c}\,. \label{spin-viel}
\end{equation} 

Nevertheless, if one considers the Yang-Mills action for the remnant Lorentz symmetry, one ends up with an $R(M)^2$ action, which is not the desired one, when aiming to reproduce the results of general relativity, because this would impose the gravitational coupling constant to be dimensionless - which is not. For the expected Einstein-Hilbert action, retrieving the correct dimensionality of the coupling constant as well, one has to consider an action in a non-straightforward way, that is building Lorentz invariants out of the quantities of the theory. The one that includes the Ricci scalar is the right choice, leading to the desired Einstein-Hilbert action.    

\section{Four-dimensional conformal gravity as a gauge theory}
Besides Einstein's gravity, discussed in section \ref{Einstein's gravity}, also Weyl's gravity is successfully formulated as a gauge theory of the conformal group, SO(2,4). Again, after the determination of the transformations of the fields and calculating the expressions of the curvature tensors, the action with which one starts is an SO(2,4) gauge invariant action of Yang-Mills type. Then, the initial gauge symmetry is broken by imposition of specific constraints on the curvature tensors, which originate from physical arguments, and the final action of the theory is the Weyl action \cite{Kaku:1977pa,Fradkin:1985am,vanproeyen}. In the present section we briefly discuss the case of the four-dimensional conformal gravity (for detailed analysis see \cite{Kaku:1977pa, vanproeyen, Fradkin:1985am}, also \cite{cham-thesis, Chamseddine:1976bf}). The gauge group parametrizing the conformal symmetry in four-dimensional Minkowski spacetime is SO(2,4), but instead, for simplicity, we consider the corresponding euclidean signature, therefore the symmetry group is the SO(6). The generators of the conformal algebra are the translations, ($P_a$), the Lorentz rotations, ($M_{ab}$), the conformal boosts, ($K_a$) and the dilatations, ($D$), satisfying the following commutation relations\footnote{Although these are the commutation relations found in literature, we have used a different but equivalent set of commutation relations for the following results, that will be given in section 4.2, equations \eqref{algebra}.}:
\begin{align}
[M_{ab},M^{cd}]&=4M_{[a}^{~[d}\delta_{b]}^{c]}\,,\quad [M_{ab},P_c]=2P_{[a}\delta_{b]c}\,,\quad [M_{ab},K_c]=2K_{[a}\delta_{b]c}\,,\nonumber \\
[P_a,D]&=P_a\,,\quad [K_a,D]=-K_a\,,\quad [P_a,K_b]=2(\delta_{ab}D-M_{ab})\,,
\end{align}  
where $a,b,c,d=1\ldots 4$. To proceed with the gauging procedure, one has to define the gauge potential:
\begin{equation}
A_{\mu} = e_{\mu}^{~a}P_{a} + \frac{1}{2}\omega_{\mu}^{~ab}M_{ab} + b_{\mu}D + f_{\mu}^{~a}K_{a}\,,\label{connection-conformal}
\end{equation}
in which for every generator, a gauge field has been corresponded. The above connection obeys the following infinitesimal transformation rule:
\begin{equation}
\delta_{\epsilon}A_{\mu} = D_{\mu}\epsilon = \partial_{\mu}\epsilon+ [A_{\mu},\epsilon]\,,\label{commrule}
\end{equation}
where $\epsilon$ is a parameter taking values in the Lie algebra of the symmetry group and therefore can be expanded on its generators:
\begin{equation}
\epsilon = \epsilon_{P}^{~a} P_{a} + \frac{1}{2}\epsilon_{M}^{~~ab} M_{ab} + \epsilon_{D}D+ \epsilon_{K}^{~a}K_{a}\,.\label{parameter-conformal}
\end{equation}
Starting from \eqref{commrule} and using  \eqref{connection-conformal} and \eqref{parameter-conformal}, calculations lead to the following transformations of the various gauge fields after comparison of the terms:
\begin{align}
\delta e_{\mu}^{~a} &= \partial_\mu\epsilon_P^{~a}+2ie_{\mu b}\epsilon_M^{~ab}-i\omega_\mu^{~ab}\epsilon_{Pb}-b_\mu\epsilon_K^{~a}+f_\mu^{~a}\epsilon_D\,, \nonumber \\
\delta \omega_{\mu}^{~ab} &= \frac{1}{2}\partial_\mu\epsilon_M^{~ab}+4ie_\mu^{~a}\epsilon_P^{~b}+\frac{i}{4}\omega_\mu^{~ac}\epsilon_{M~c}^{~~b}+if_\mu^{~a}\epsilon_K^{~b}\,, \nonumber \\
\delta b_{\mu} &= \partial_\mu\epsilon_D-e_\mu^{~a}\epsilon_{Ka}+f_\mu^{~a}\epsilon_{Pa}\,, \nonumber \\
\delta f_{\mu}^{~a} &= \partial_\mu\epsilon_K^{~a}+4ie_\mu^{~a}\epsilon_D-i\omega_\mu^{~ab}\epsilon_{Kb}-4ib_\mu\epsilon_P^{~a}+if_\mu^{~b}\epsilon_{M~b}^{~~a}\,. \label{conformaltrans}
\end{align}
The well-known formula for the curvature:
\begin{equation}
R_{\mu\nu} = 2 \partial_{[\mu} A_{\nu]} - i [ A_{\mu} ,A_{\nu} ]
\end{equation}
gives the expressions of the component curvatures of the gauge fields of the theory:
\begin{align}
R_{\mu\nu}^{~~~a}(P) &= 2 \partial_{[\mu}e_{\nu]}^{~~a} + f_{[\mu}^{~~a}b_{\nu]} +  e^{~~b}_{[\mu} \omega_{\nu]}^{~~ac} \delta_{bc}, \nonumber \\
R_{\mu\nu}^{~~~ab}(M) &= \partial_{[\mu} \omega_{\nu]}^{~~ab} + \omega_{[\mu}^{~~ca} \omega_{\nu]}^{~~db} \delta_{cd} + e_{[\mu}^{~~a}e_{\nu]}^{~~b} + f_{[\mu}^{~~a}f_{\nu]}^{~~b}, \nonumber \\
R_{\mu\nu}(D) &= 2 \partial_{[\mu}b_{\nu]} + f_{[\mu}^{~~a}e_{\nu]}^{~~b}\delta_{ab}, \nonumber \\
R_{\mu\nu}^{~~~a}(K) &= 2 \partial_{[\mu}f_{\nu]}^{~~a} + e_{[\mu}^{~~a}b_{\nu]} + f_{[\mu}^{~~b}\omega_{\nu]}^{~~ac}\delta_{bc}\,. \label{conformalcurvatures}
\end{align}
The action is taken to be of Yang-Mills form, being SO(6) gauge invariant. According to \cite{Fradkin:1985am, vanproeyen,Kaku:1977pa}, there exist some convincing arguments that give specific constraints which induce the symmetry breaking of the initial symmetry. In particular, these constraints are the torsionless condition, $R(P)=0$ along with an additional constraint on $R(M)$, together leading to expressions of $\omega_\mu^{~ab}$ and $f_\mu^{~a}$ in terms of $e_\mu^{~a}$ and $b_\mu$ and finally, as a third constraint, the fact that $b_\mu$ gauge field can be gauged away. The resulting action is the well-known Weyl action. 

Besides the above arguments, we suggest another way of breaking the initial symmetry, this time to the Lorentz. This breaking, along with the transformations of the fields and component tensor expressions we presented above, serves our purpose for the present paper, because we want to use these results as the commutative limit of the noncommutative gravity theory we develop later. This could occur with the inclusion of two scalars in the $\mathrm{6}$ representation of the SO(6) gauge group \cite{Li:1973mq}. This comes in a very natural way, taking one step beyond the way the de Sitter group (and not the Poincar\'e for reasons explained in the previous section) is broken to the Lorentz by a scalar in the fundamental of SO(1,4), as described in the previous section for the case of the Einstein gravity. These two scalars could induce a spontaneous symmetry breaking in a complete theory that includes matter fields, giving rise to the constraints that lead us to a resulting four-dimensional action that respects Lorentz symmetry. 

Moreover, we can also extend the argument we used in the previous section, in the four-dimensional Poincar\'e gravity, as an alternative way to break the initial symmetry. Since we want to preserve the Lorentz symmetry out of the initial SO(6), we consider directly that the vacuum of the theory is SO(4) invariant, which in turn means that every other tensor, except for the $R(M)$, has to vanish. This vanishing will produce the constraints of the theory leading to specific expressions that relate the gauge fields. In particular, in \cite{Chamseddine:2002fd}, it is described that if both tensors $R(P)$ and $R(K)$ are set simultaneously equal to zero, then the constraints of the theory yield that their corresponding gauge fields, $f_\mu^{~a}, e_\mu^{~a}$ are equal - up to a rescaling factor - and also $b_\mu=0$.
 
\section{Four-dimensional noncommutative gravity on a fuzzy covariant space} 

\subsection{Construction of the fuzzy covariant space}

In this subsection, we present the construction of the four-dimensional fuzzy space, on which we elaborate gravity as a gauge theory. Although, generally in gravity theories, the spacetime is obtained as a solution of the equations of motion, in the gauge-theoretic approach one has to predicate the background spacetime in order to extract the information of the isometry group which is eventually gauged. The corresponding gauge theory is developed on this space.  

We want to construct a gravitational model on a four-dimensional fuzzy space, specifically the fuzzy version of the $\text{dS}_4$. The de Sitter space is defined as a submanifold of the five-dimensional Minkowski spacetime in the same way the four-sphere is defined as an embedding in the five-dimensional euclidean space. Specifically, the embedding equation that defines $\text{dS}_4$ is:
\begin{equation}
\eta^{AB}x_Ax_B=R^2\,,
\end{equation}
where $A,B=0,\ldots, 4$ and $\eta^{AB}$ is the mostly positive metric of the five-dimensional Minkowski spacetime, that is $\eta^{AB}=\text{diag}(-1,+1,+1,+1,+1)$.

However, the coordinates of the fuzzy de Sitter space, $X_a$, must satisfy the following commutation relation, which describes the noncommutativity of the space:
\begin{equation}
[X_a,X_b]=i\theta_{ab}\,.\label{noncommgeneral}
\end{equation}
Recalling the case of the fuzzy two-sphere with coordinates the three rescaled SU(2) generators in an (large) N-dimensional representation and radius $r$, the right hand side of \eqref{noncommgeneral} is also a generator of the SU(2) algebra, ensuring covariance, that is $\theta_{ab}={C}_{abc}{X}_c$, where $C_{abc}$ is a rescaled Levi-Civita symbol. In the fuzzy de Sitter space we want to construct, the problem is that when trying to identify the coordinates with some generators of SO(1,4), covariance is broken, because the algebra is not closing, i.e. $\theta_{ab}$ cannot be assigned to generators into the algebra \cite{Heckman:2014xha}\footnote{For a detailed analysis of this issue, see also \cite{Kimura:2002nq,Sperling:2017dts}, where the authors encounter the same problem when trying to build the fuzzy four-sphere.}. The requirement of the preservation of covariance, imposes us to use a group with a larger symmetry, in which we will be able to incorporate all generators and the noncommutativity in it. The minimum extension of the symmetry leads to the adoption of the SO(1,5) group. Therefore, in order to end up with a fuzzy $\text{dS}_4$, with coordinates represented by N-dimensional matrices (like in fuzzy two-sphere case), respecting covariance as well, we are led to the enlargement of the symmetry to the SO(1,5) group. From now on, for convenience, we will use the euclidean signature, which means that the isometry group and the extended symmetry group will be SO(5) and SO(6), respectively, instead of SO(1,4) and SO(1,5)\footnote{This translation to the euclidean signature could be interpreted as the case of the fuzzy four-sphere which bears the same isometry group SO(5), along with the  restriction that large-N matrices are set into irreducible representations, which implies the imposition of the radius constraint.}.   
  
Now, in order to formulate the above four-dimensional fuzzy space, we consider the extended algebra of SO(6). We denote its generators by $J_{AB} = - J_{BA}$, with $A,B = 1,\ldots, 6$ and they satisfy the following commutation relations:
\begin{equation}
[J_{AB}, J_{CD} ] = i(\delta_{AC}J_{BD} + \delta_{BD}J_{AC} - \delta_{BC}J_{AD} - \delta_{AD}J_{BC} )\,.
\end{equation}
Now, we decompose the above generators in an SO(4) notation\footnote{The embedding path we consider is $SO(6) \supset SO(5) \supset SO(4)$.}, identifying the component generators as: 
\begin{equation}
J_{mn} = \frac{1}{\hbar} \Theta_{mn}, \ \ J_{m5} = \frac{1}{\lambda} X_{m}, \ \  J_{m6} = \frac{\lambda}{2 \hbar}P_{m} , \ \  J_{56} = \frac{1}{2} \mathrm{h}\,,
\end{equation}
with $m = 1,\ldots,4$. For dimensional reasons we have introduced an elementary length $\lambda$. The coordinates, momenta and non-commutativity  tensors are $X_{m}$, $P_{m}$ and $\Theta_{mn}$, respectively.
Both coordinates and momenta satisfy the following algebra:
\begin{equation}
[ X_{m} , X_{n} ] = i \frac{\lambda^{2}}{\hbar} \Theta_{mn}, \ \ \ [P_{m}, P_{n} ] = 4i \frac{\hbar}{\lambda^{2}} \Theta_{mn}
\end{equation}
\begin{equation}
[ X_{m}, P_{n} ]  = i \hbar \delta_{mn}\mathrm{h}, \ \ \ [X_{m}, \mathrm{h} ] = i \frac{\lambda^{2}}{\hbar} P_{m}
\end{equation}
\begin{equation}
[P_{m}, \mathrm{h} ] =4i \frac{\hbar}{\lambda^{2}} X_{m}\,,
\end{equation}
where $m,n = 1,\dots, 4$. The extended kinematical algebra i.e. the algebra of spacetime transformations is:
\begin{equation}
[X_{m}, \Theta_{np} ] = i \hbar ( \delta_{mp} X_{n} - \delta_{mn} X_{p} )
\end{equation}
\begin{equation}
[P_{m}, \Theta_{np} ] = i \hbar ( \delta_{mp} P_{n} - \delta_{mn} P_{p} )
\end{equation}
\begin{equation}
[\Theta_{mn}, \Theta_{pq} ] = i \hbar ( \delta_{mp} \Theta_{nq} + \delta_{nq} \Theta_{mp} - \delta_{np} \Theta_{mq} - \delta_{mq} \Theta_{np} )
\end{equation}
\begin{equation}
[\mathrm{h}, \Theta_{mn} ] = 0
\end{equation}
The above algebra, in contrast to the Heisenberg algebra (see \cite{Singh:2018qzk}), admits finite dimensional representations for $X_{m}$, $P_{m}$ and $\Theta_{mn}$, thus we have obtained a model of spacetime which is a finite quantum system. In analogy with the fuzzy two-sphere, spaces like the one above are called fuzzy covariant non-commutative spaces \cite{Barut, Heckman:2014xha,Buric:2015wta}. In the following section we will formulate a gauge theory of gravity on the above four-dimensional fuzzy space.

\subsection{Noncommutative gauge theory of four-dimensional gravity}

Here, we are going to present the formulation of gravity as a gauge theory on the four-dimensional space that was constructed in the previous section\footnote{For a noncommutative gravity motivated from string theory see \cite{AlvarezGaume:2006bn}.}. The whole procedure is developed as a noncommutative analogue of the work presented in sections 2 and 3, using upon them the framework of noncommutative gauge theories \cite{Madore:2000en}. \\

\noindent\emph{Determination of the gauge group and representation by $4 \times 4$ matrices}
\\\\
\noindent Recalling the previous section, we argued that for the sake of preserving gauge covariance, we were led to enlarge the symmetry group of the space to the SO(6). Thinking along the lines of the commutative four-dimensional gravity, described in section 2, where the isometry group, (the Poincar\'e group), was gauged in order to retrieve the desired results, here we have to gauge the isometry group of the covariant space, that is the SO(5), as seen as a subgroup of the SO(6), in which we ended up with.  

However, it is known that in noncommutative gauge theories, the use of the anticommutators of the generators of the algebra is inevitable, as we have explained in detail in our previous works \cite{Chatzistavrakidis:2018vfi,Manolakos:2018hvn,Manolakos:2018isw} (see also \cite{Aschieri:2009ky}). Of course, the anticommutators of the generators (not in a specific representation) of an algebra do not necessarily yield operators that belong to the algebra and this is exactly the case for the generators of SO(5). The indicated treatment is to specify the representation in which the generators belong and include the operators produced by the anticommutators into the algebra, considering them as generators, too. This will result in the extension of the initial gauge group to one with larger symmetry. In our case, application of this recipe leads to the extension of SO(5) to the SO(6)$\times$U(1) ($\sim$U(4)) group\footnote{The extension to SO(6)$\times$U(1) is a coincidence with the SO(6) symmetry related to our space and should not be confused.} with generators being represented by 4x4 matrices, i.e. the representation should be fixed to the $\mathrm{4}$  of SO(6) i.e. to the fundamental representation of U(4).

We start with the four $\gamma$-matrices (in the Euclidean signature) satisfying:
\begin{equation}
\{\Gamma_{a}, \Gamma_{b} \} = 2 \delta_{ab} \one\,,
\end{equation} 
where $m,n = 1,\ldots 4$ and also defining $\Gamma_{5}$ as $ \Gamma_{5} = \Gamma_{1} \Gamma_{2} \Gamma_{3} \Gamma_{4} $. The generators of the SO(6) group are:\\\\
a) Six Lorentz rotation generators:
$ \mathrm{M}_{ab} =  - \dfrac{i}{4} [\Gamma_{a} , \Gamma_{b} ] = - \dfrac{i}{2} \Gamma_{a} \Gamma_{b}\,,a < b$,\\\\
b) four generators for conformal boosts: $ \mathrm{K}_{a} = \dfrac{1}{2} \Gamma_{a}$,\\\\
c) four generators for translations: $ \mathrm{P}_{a} = -\dfrac{i}{2} \Gamma_{a} \Gamma_{5}$,\\\\
d) one generator for special conformal transformations: $\mathrm{D} = -\dfrac{1}{2} \Gamma_{5}$ and\\\\
e) one $\mathrm{U(1)}$ generator: $ \one$. \\\\ 
The construction of $\Gamma$-matrices starts with Pauli matrices:
\begin{equation}
\sigma_{1} =\left(
            \begin{array}{cc}
            0 & 1 \\
            1 & 0 \\
            \end{array}
            \right)    \ \ \ \ \ \
\sigma_{2} = \left(
               \begin{array}{cc}
                 0 & -i \\
                 i & 0 \\
               \end{array}
             \right)    \ \ \ \ \ \ \
\sigma_{3} = \left(
               \begin{array}{cc}
                 1& 0 \\
                 0 & -1 \\
               \end{array}
             \right)
\end{equation}
and $\gamma-$matrices are built as tensor products of them as:
$$ \Gamma_{1} = \sigma_{1} \otimes \sigma_{1}, \ \ \ \Gamma_{2} = \sigma_{1} \otimes \sigma_{2}, \ \ \ \Gamma_{3} = \sigma_{1} \otimes \sigma_{3} $$
$$ \Gamma_{4} = \sigma_{2} \otimes \mathbf{1}, \ \ \  \Gamma_{5} = \sigma_{3} \otimes \mathbf{1}\,. $$
In particular, we obtain:
\begin{equation}
M_{ij} = - \frac{i}{2}\Gamma_{i} \Gamma_{j} = \frac{1}{2} \mathbf{1} \otimes \sigma_{k}\,,
\end{equation}
where $i,j,k = 1,2,3$ and:
\begin{equation}
M_{4k} = - \frac{i}{2}\Gamma_{4} \Gamma_{k} = - \frac{1}{2} \sigma_{3} \otimes \sigma_{k}
\end{equation}

The generators defined above, obey the following algebra:
\begin{eqnarray}
&&[ K_{a} , K_{b} ] = i M_{ab}, \ \ \ [P_{a}, P_{b} ] = i M_{ab} \nonumber \\
&&[P_{a}, D ] =i K_{a} , \ \ \ [K_a,P_b]=i\delta_{ab}D , \ \ \ [K_a,D]=-iP_a \nonumber \\
&&[K_{a}, M_{bc} ] = i( \delta_{ac} K_{b} - \delta_{ab} K_{c} ) \nonumber \\
&&[P_{a}, M_{bc} ] = i( \delta_{ac} P_{b} - \delta_{ab} P_{c} ) \nonumber \\
&&[M_{ab}, M_{cd} ] = i( \delta_{ac} M_{bd} + \delta_{bd} M_{ac} - \delta_{bc}M_{ad} - \delta_{ad}M_{bc} ) \nonumber \\
&&[D, M_{ab} ] = 0\,.\label{algebra}
\end{eqnarray}

\noindent\emph{Noncommutative gauge theory of gravity}
\\\\
Previously, we decomposed the SO(6) generators of the SO(6)$\times$U(1) in the SO(4) notation, in order to identify the generators in the appropriate language and then we calculated their commutation relations. Therefore, we are ready to advance with the (noncommutative) gauging procedure. 

First, one has to define the covariant coordinate of the theory, which is given by the following relation:
\begin{equation}
\hat{X}_m=X_m\otimes\one+A_m(X)\,.\label{covcoordinatedefinition}
\end{equation}
The coordinate $\hat{X}_m$ is covariant by construction, which means that it transforms covariantly under a gauge transformation:
\begin{equation}
\delta\hat{X}_m=i[\epsilon,\hat{X}_m]\,,\label{covcoord}
\end{equation}
where $\epsilon=\epsilon(X)$ is the gauge parameter, which is a function of the coordinates, but also takes values in the algebra, \eqref{algebra}, given in the previous section. Therefore, it can be written as an expansion on the generators of the algebra. Consequently, the gauge parameter is written explicitly as:
\begin{equation}
\epsilon(X)=\epsilon_0(X)\otimes\one+\xi^a(X)\otimes K_a+\tilde{\epsilon}_0(X)\otimes D+\lambda^{ab}(X)\otimes M_{ab}+\tilde{\xi}^a(X)\otimes P_a\,.\label{parameterso(4)}
\end{equation}
Considering that the gauge transformation does not affect the coordinate $X_m$, i.e. $\delta X_m=0$, one can find the transformation property of the $A_m$ included in the \eqref{covcoord}. In analogy with the commutative case, the way $A_m$ transforms indicates that it can be considered as the potential, that is the gauge connection of the theory. In our case, $A_m$ is a function of matrices-coordinates $X_a$ of the fuzzy $\mathrm{dS_4}$. The $A_m(X)$ takes values in the U(4) algebra, meaning that it can be spanned on its generators:
\begin{equation}
A_m(X)=e_m^{~a}(X)\otimes P_a+\omega_{m}^{~ab}(X) \otimes  M_{ab}(X) + b_{m}^{~a}(X) \otimes K_{a}(X) + \tilde{a}_{m}(X) \otimes D + a_{m}(X) \otimes  \one\,.
\end{equation}  
From the above equation, one can read that we have introduced one gauge field for each generator. The component gauge fields depend on the coordinates of the space, $X_a$, meaning that they are N$\times$N matrices, where N is the dimension of the representation in which the coordinates are described. Multiplication between every gauge field and its corresponding generator is not the usual one, but the tensor product is used, since the product consists of matrices of different dimensions, recalling that the generators are described by 4$\times$4 matrices. Therefore, every term of the connection is a 4N$\times$4N matrix. 

\noindent Having determined the gauge connection, the covariant coordinate is written explicitly as:
\begin{equation}
\hat{X}_{m} = X_{m} \otimes \one + e_{m}^{~a}(X) \otimes P_{a} + \omega_{m}^{~ab}(X) \otimes M_{ab} + b_{m}^{~a} \otimes K_{a} + \tilde{a}_{m} \otimes D + a_{m} \otimes  \one\,.\label{covcoordinatedecomposition}
\end{equation}
Furthermore, for the U(4) gauge theory we are developing, what is left to determine is the field strength tensor\footnote{Details on its definition and gauge covariance property are given in Appendix A.}, which, for the noncommutative case, is defined as:
\begin{equation}
\mathcal{R}_{mn} = [\hat{X}_{m}, \hat{X}_{n}]  - \frac{i\lambda^2}{\hbar}\hat{\Theta}_{mn}\,,\label{fieldstrengtt}
\end{equation}
which, since it is valued in the algebra, it is expanded in terms of the component curvatures as:
\begin{equation}
\mathcal{R}_{mn}(X) = R_{mn}^{~~~ab}(X) \otimes M_{ab} + \tilde{R}_{mn}^{~~~a}(X) \otimes P_{a} + R_{mn}^{~~~a}(X) \otimes K_{a} + \tilde{R}_{mn}(X) \otimes D + R_{mn}(X) \otimes \one\,.\label{fieldstrengthtensordecomposition}
\end{equation}
At this point, all that is necessary for the determination of the transformations of the gauge fields and the expressions of the component curvatures is obtained. The results and some intermediate calculations are given in the Appendix B.

\section{The constraints for breaking the symmetry to SO(4)$\times$U(1)}

%\noindent\emph{The constraints for breaking the symmetry}
%\\\\
The gauge symmetry we would like to end up with is the one described by the Lorentz group, in the euclidean signature we use, the SO(4). A straightforward way to do this is by considering a constrained theory in which the rest of the component tensors are vanishing, in order to break the initial SO(6)$\times$U(1) symmetry to the desired SO(4)$\times$U(1), which will be the remnant gauge symmetry of the vacuum of the field theory. However, this approach leads to an over-constrained theory, that is evident after counting the degrees of freedom that survive after the breaking. Therefore, it is rather wise to impose at first the constraint:
\begin{equation}
\tilde{R}_{mn}^{~~~a}(P)=0\,, \label{ncconstraints}
\end{equation}
that is the torsionless condition, which is more or less expected, as it is considered in previous sections as well, in the cases of Einstein and conformal gravity.
Furthermore, the possible interpretation of $b_m^{~a}$ as a second vielbein\footnote{We note that our symmetry group is not the conformal group but is related to it.} would give a theory with two metrics or two vielbeins, which is not desirable in our case. Therefore, we are led to adopt as a solution of the constraints, the $e_m^{~a}=b_m^{~a}$. The consideration of this solution leads to the expression of the spin connection $\omega_m^{~ab}$ as a function of the rest of the fields, ${e}_m^{~a}, {a}_m, {\tilde{a}}_m$. Taking into consideration the torsionless condition with the specific solution $e_m^{~a}=b_m^{~a}$, we result with the correct number of degrees of freedom, leading to the noncommutative analogue of the four-dimensional gravity with reduced symmetry SO(4)$\times$U(1). 

\noindent In order to go on with obtaining the expression relating the fields, we employ the following two identities:
\begin{equation}
\delta^{abc}_{fgh}=\epsilon^{abcd}\epsilon_{fghd}\quad\quad\text{and}\quad\quad \frac{1}{3!}\delta^{abc}_{fgh}a^{fgh}=a^{[fgh]}\,.\label{ids}
\end{equation} 
The constraint of the torsionless condition takes the following form:
\begin{equation}
\epsilon^{abcd}[e_m^{~b},\omega_n^{~cd}]-i\{\omega_m^{~ab},e_{nb}\}=-[D_m,e_m^{~a}]-i\{e_m^{~a},\tilde{a}_m\}\,,
\end{equation}
where $D_m=X_m+a_m$, i.e. the covariant coordinate of an abelian gauge theory. The above equation leads to the following two:
\begin{equation}
\epsilon^{abcd}[e_m^{~b},\omega_n^{~cd}]=-[D_m,e_m^{~a}]\quad\quad \text{and}\quad\quad \{\omega_m^{~ab},e_{nb}\}=\{e_m^{~a},\tilde{a}_n\}\,.
\end{equation}
Using the identities \eqref{ids}, the above equations lead to the desired expression of the spin connection in terms of the rest of the fields:
\begin{equation}
\omega_n^{~ac}=-\frac{3}{4}e^m_{~b}(-\epsilon^{abcd}[D_m,e_{nd}]+\delta^{[bc}\{e_n^{~a]},\tilde{a}_m\})\,.\label{omegaintermsofe}
\end{equation}

According to \cite{Green:1987mn}, the vanishing of the field strength tensors in a gauge theory considered on a simply connected space means that, locally, the corresponding gauge fields vanish as well. This argument would be very welcome in our case because it would certainly simplify the expressions of the tensors. Nevertheless, the above argument cannot be applied in our case in general, because after identifying the vielbein with gauge fields of our spacetime, we mixed gauge theory and geometry. It holds that the vielbein is considered to be invertible in every point, therefore, adopting the above argument, that is setting the vielbein to zero, would lead to degenerate vielbein matrices, inducing degeneracy to the metric tensor of the space \cite{Witten:1988hc}. However, we could set $\tilde{a}_m=0$. This will also modify the \eqref{omegaintermsofe}, producing an even simpler expression of the spin connection in terms of the vielbein, $e_m^{~a}$, and the $a_m$ (the two surviving fields of the theory), which reads:
\begin{equation}
\omega_n^{~ac}=\frac{3}{4}e^m_{~b}\epsilon^{abcd}[D_m,e_{nd}]\,.\label{omegaintermsofeteliko}
\end{equation}     

At this point we need to punctuate that the U(1) field strength tensor, $R_{mn}(\one)$ is not set to zero, which means that this U(1), strongly related to the noncommutativity, remains unbroken in the resulting theory after the breaking, since we still have a theory on a noncommutative space. Of course, the corresponding field, $a_m$, would vanish if we assumed the commutative limit of the broken theory, in which noncommutativity is lifted and ${a_m}$ decouples being super-heavy. In this limit, the gauge theory would be just SO(4).    

Another way of getting the desired SO(4) symmetry as a remnant symmetry after the breaking of SO(6), is to extrapolate the argument that we developed for the conformal gravity in the commutative case (section 3) of the inclusion of two scalar fields in the fundamental representation of SO(6), to the noncommutative case. We are confident that the spontaneous symmetry breaking will lead to constraints equivalent to the one in \eqref{ncconstraints}. \\

%\noindent\emph{Gauge covariant general coordinate transformations and constraints}\\\\
%\noindent We can also define the gauge covariant general coordinate transformations, which have the desired commutative limit, as:
%\begin{equation}
%\delta_C(f)A_n^{~A}=\frac{1}{2}\{f^m,R_{mn}^{~~~A}\}+\frac{1}{2}\delta_G(\{f^m,A_m^{~A}\})A_n^{~A}\,,
%\end{equation}
%where the index $C$ and the parameter $f$ correspond to an infinitesimal coordinate transformation, while the index $G$ and the parameter $\epsilon^A=\dfrac{1}{2}\{f^m,A_m^{~A}\}$ correspond to a gauge transformation. A parameter $\epsilon^a$ related to gauge transformations generated by $P, K$ can be constructed using the gauge fields $e_m^{~a}, b_m^{~a}$ as $\epsilon^a=f^me_m^{~a}$ and $\epsilon^a=f^mb_m^{~a}$. The covariant coordinate transformations of the fields $e_m^{~a}$ and $b_m^{~a}$ are:
%\begin{equation}
%\delta_C(f)e_n^{~a}=\frac{1}{2}\{f^m,\tilde{R}_{mn}^{~~~a}\}+\delta_P(\epsilon^a)e_n^{~a}\quad\text{and}\quad \delta_C(f)b_n^{~a}=\frac{1}{2}\{f^m,R_{mn}^{~~~a}\}+\delta_K(\epsilon^a)b_n^{~a}\,.
%\end{equation}
%We observe that a gauge transformation generated by $P$ and $K$ corresponds to a coordinate transformation, if and only if $\tilde{R}_{mn}^{~~~a}=0$ and $R_{mn}^{~~~a}=0$.
\section{The action}

\noindent Since we have employed a gauge theory formulation for constructing the gravitational model, the most reasonable choice for the action is of Yang-Mills  form\footnote{A Yang-Mills action $\text{tr}F^2$ is gauge invariant, as it is explained in Appendix A.}:
\begin{equation}
\mathcal{S}=\text{Tr} \, \text{tr} \, \Gamma_5 \, \left(\mathcal{R}_{mn}\mathcal{R}_{rs}\epsilon^{mnrs}+\hat{\mathcal{H}}_{mnp}\hat{\mathcal{H}}^{mnp}\right)\,,\label{action48}
\end{equation}
where the $\text{Tr}$ is the trace over the matrices representing the coordinates (takes the role of the integration of the commutative case) whereas the $\text{tr}$ is the trace over the generators of the algebra. The first term of the above action includes the field strength (curvature) tensor of the gauge theory, while the second one is the (non-topological) kinetic term of the 2-form field. Writing down the decompositions of the tensors in terms of their component tensors, \eqref{fieldstrengthtensordecomposition}, \eqref{fieldstrengthHdecomposition} and applying the $\text{tr}$ on the generators, we end up with the following expression of the action:
\begin{equation}
\mathcal{S}=\text{Tr}\left(\frac{\sqrt{2}}{4}R_{mn}^{~~~ab}R_{rs}^{~~cd}\epsilon_{abcd}-4\tilde{R}_{mn}R_{rs}\right)\epsilon^{mnrs}+\text{Tr}\left(\frac{\sqrt{2}}{4}H_{mnp}^{~~~~ab}H^{mnpcd}\epsilon_{abcd}-4\tilde{H}_{mnp}H^{mnp}\right),\label{actionaftertrace}
\end{equation}
where in order to result with the above expression, the commutation and anticommutation relations \eqref{algebra}, \eqref{anticomso(4)} for the generators of the algebra have been used, as well as the cyclicity property of the trace.

\noindent Considering the vacuum state of the theory, i.e. applying all the constraints considered in the previous section:
\begin{equation}
    e_m^{~a}=b_m^{~a}\,,\quad \tilde{a}_m=0\,,\quad \omega_m^{~ab}=\frac{3}{4}e^n_{~c}\epsilon^{abcd}[D_m,e_{nd}]
\end{equation}
and inserting the expressions of the component tensors given in \eqref{R(1)}-\eqref{R(M)} (modified according to the above constraints) in the above action, will give the explicit expression of the action after the symmetry breaking:
\begin{align}
\mathcal{S}&=\frac{\sqrt{2}}{4}\text{Tr}\left(\frac{3}{4}[D_m,e^p_{~d}\epsilon^{abef}[D_m,e_{pf}]]-\frac{3}{4}[D_n,e^p_{~e}\epsilon^{abef}[D_n,e_{pf}]]+i\{e_m^{~a},e_n^{~b}\}\right.\nonumber\\
&~~~~~~~~~~~~~+\left.\frac{9i}{8}\{e^p_{~d}\epsilon^{acde}[D_m,e_{pe}],e^q_{~f}\epsilon^{b~fg}_{~c}[D_n,e_{qg}]\}-\frac{i\lambda^2}{\hbar}B_{mn}^{~~~ab}\right)\cdot\nonumber\\
&~~~~~~~~~~~~\left(\frac{3}{4}[D_r,e^p_{~e}\epsilon^{cdef}[D_s,e_{pf}]]-\frac{3}{4}[D_s,e^p_{~e}\epsilon^{cdef}[D_r,e_{pf}]]+i\{e_r^{~c},e_s^{~d}\}\right.\nonumber\\
&~~~~~~~~~~~~~+\left.\frac{9i}{8}\{e^p_{~j}\epsilon^{chje}[D_r,e_{pe}],e^q_{~f}\epsilon^{d~fg}_{~h}[D_s,e_{qg}]\}-\frac{i\lambda^2}{\hbar}B_{rs}^{~~~cd}\right)\epsilon^{mnrs}\nonumber\\
    -4\text{Tr}&\left(-\frac{9\sqrt{2}}{16}\epsilon^{efgh}[e^p_{~c}[D_m,e_{pf}],e^q_{~g}[D_n,e_{qh}]]-\frac{i\lambda^2}{\hbar}\tilde{B}_{mn}\right)\epsilon^{mnrs}\cdot\nonumber\\
    &\left([D_r,D_s]+\frac{5}{16}[e_r^{~a},e_{sa}]+\frac{9}{32}[e^q_{~c}[D_r,e_{qd}],e^{pc}[D_s,e_p^{~d}]-e^{pd}[D_s,e_p^{~c}]]-\frac{i\lambda^2}{\hbar}(\Theta_{rs}+B_{rs})\right)\nonumber\\
   &\hspace{-1cm}+\text{Tr}\left(\frac{\sqrt{2}}{4}H_{mnp}^{~~~~ab}H^{mnpcd}\epsilon_{abcd}-4\tilde{H}_{mnp}H^{mnp}\right)\,.\label{actionftertheconstraints}
\end{align}
It should be noted that a gauge transformation on the above action (after the symmetry breaking) is expected to leave it invariant. This could be proved by making use of the explicit expressions of the gauge transformations of the component tensors, which are given in \eqref{transforamtionsofthecomponenttensots} in Appendix B\footnote{Explicit calculations are quite cumbersome to prove the gauge invariance of the action in a straightforward way. Nevertheless, based on the corresponding commutative cases which inspired our choices for the symmetry breaking, we are quite confident that gauge invariance should be secured. Also, it is postponed for our future work to include scalar fields in the theory and realize the breaking of the initial symmetry spontaneously. This approach, among other welcome features, will directly ensure the gauge invariance of the produced action after the symmetry breaking.}. Last, variation with respect to the (surviving) gauge fields would lead to the equations of motion. 

\section{Summary and Conclusions}

In the present work we have presented a four-dimensional gravity model on a covariant noncommutative space. The chosen space is a generalization of the celebrated fuzzy two-sphere in certain basic aspects and is a noncommutative version of $\mathrm{dS_4}$, keeping the fundamental property of covariance of the two-sphere, which preserves all isometries of the fuzzy space and, in particular, the Lorentz invariance, which is of particular interest in the present four-dimensional case. Another property that shares with the fuzzy two-sphere is that its coordinates can be represented by finite matrices. Then gravity is built by gauging the isometries of the constructed noncommutative space. It should be noted though that the requirement of covariance led us to an enlargement of the isometries of the fuzzy $\text{dS}_4$, while, furthermore, the construction of a gauge theory on such a noncommutative space led us to an enlargement of the gauge group and in fixing its representation. Also, the definition of the field strength tensor imposed the inclusion of a 2-form dynamic gauge field for reasons of covariance. The introduction of such a gauge field in gauge theories constructed on covariant spaces like the one we use, is a treatment that can be applied in general. That is why we described in detail this subject (Appendix A), for bookkeeping and use as a future reference. 

Then, straightforward calculations led to the transformations of the gauge fields and the expressions of the component tensors. However, the part of the isometry group we decided to gauge, even more after the enlargement noncommutativity imposed, was larger than the one we wanted to result with. For this reason, we considered a symmetry breaking which led to a constrained gauge theory, with the preferred symmetry. In this spirit, we proposed an SO(6)$\times$U(1) action of Yang-Mills type, including the kinetic term of the dynamic 2-form field. Imposition of the constraints leads to an expression of the action which can be varied in order to obtain the equations of motion. Note that in the commutative limit we recover the conformal gravitational theory, not the one of Einstein's general relativity.  

The noncommutative gravity model we propose hopefully provides a certain insight concerning the gravitational interaction at the Planck scale, in which noncommutativity is naturally introduced. In addition, the fact that the commutative limit of our model leads to an already existing theory (conformal gravity) is a very welcome result. Therefore, the picture that emerges from our construction is that at high energies the coordinates are considered to be noncommutative and are also part of the commutation relations of the conformal group. The low-energy limit can be obtained by Wigner-In\"on\"u contraction, leading to the usual commutation relations of quantum mechanics \cite{Kastrup:1966zzb}. In addition, our work consists a formulation of a four-dimensional covariant fuzzy space, providing a gravity realization of the starting suggestions of Heckman-Verlinde \cite{Heckman:2014xha}.

In addition, let us make a couple of comments on our approach. First, we should note that we did not aim at a quantization (in the usual sense) of the gravity theory we proposed. For this reason the vielbein field was considered to be invertible at every point of the space and, under this consideration, singularities related to this issue were automatically avoided \cite{Witten:1988hc}. Also, it should be noted that our present noncommutative approach is not related to Ashtekar's formalism or loop quantum gravity in general.  

Finally, it should be stressed that the constructed gravity model is a matrix model giving promises for improved UV properties as compared to ordinary gravity. Clearly, the latter, as well the inclusion of matter fields is going to be a subject of further study.

\appendix
\appendixpage
%\addappheadtotoc
\section{Gauge covariant field strength tensor and gauge invariance}

In this appendix, we briefly present the gauge covariance of the field strength tensor and the gauge invariance of the action we propose. 

In general, the field strength tensor of a noncommutative gauge theory can be written in terms of the covariant coordinate plus an extra term which contributes in such a way that the whole expression of the field strength tensor transforms covariantly under a gauge transformation. The two most widely-known types of spaces on which gauge theories are built are those of constant and Lie-type noncommutativity, specifically defined by the relations: 
\begin{equation}
[X_m,X_n]=i\theta_{mn}\,,\quad\quad [X_m,X_n]=iC_{mnp}X_p \,,
\end{equation}
where $\theta_{mn}$ is a constant antisymmetric tensor and $C_{mnp}$ is a rescaled Levi-Civita symbol. Let $\hat{X}_m$ be the covariant coordinate in both cases, then expressions of the field strength tensors are given by:
\begin{equation}
F_{mn}=[\hat{X}_m,\hat{X}_n]-i\theta_{mn}\,,\quad\quad F_{mn}=[\hat{X}_m,\hat{X}_n]-iC_{mnp}\hat{X}_p\,.
\end{equation}

On the one hand, the fuzzy space on which we are building the gauge theory of gravity is covariant, pointing at a relation with the fuzzy sphere case, but, on the other hand, the right hand side of the relation defining non-commutativity does not depend on the coordinates, pointing at the case of constant noncommutativity. The truth is that it cannot be classified into any of these cases, therefore it has to be examined explicitly.

The commutation relation of the fuzzy space in our case is:
\begin{equation}
[X_m,X_n]=i\frac{\lambda^2}{\hbar}\Theta_{mn}\otimes\one\,.
\end{equation}
Due to the fact that the right hand side does not depend on the coordinates $X_m$, one would be led to a definition of the corresponding field strength tensor as:
\begin{equation}
{F}_{mn} = [\hat{X}_{m}, \hat{X}_{n}]  - \frac{i\lambda^2}{\hbar}\Theta_{mn}\otimes\one\,.
\end{equation} 
Nevertheless, if we assume a gauge transformation\footnote{A gauge transformation is considered not to affect the coordinates, that is $\delta X_m=0$ and consequently $\delta\Theta_{mn}=0$.} of the field strength tensor, straightforward calculations lead to a result according to which, the field strength tensor does not transform covariantly, specifically:
\begin{equation}
\delta F_{mn}=[\epsilon,F_{mn}]-\frac{i\lambda^2}{\hbar}[\epsilon,\Theta_{mn}\otimes\one]\,,
\end{equation} 
where $\epsilon$ is a gauge parameter. In order to ameliorate the above problematic result, one has to modify the definition of the field strength tensor to:
\begin{equation}
\hat{F}_{mn}=[\hat{X}_m,\hat{X}_n]-\frac{i\lambda^2}{\hbar}\hat{\Theta}_{mn}\,,\label{hatfieldstrength}
\end{equation}
where $\hat{\Theta}_{mn}=\Theta_{mn}\otimes\one+\mathcal{B}_{mn}$, where $\mathcal{B}_{mn}$ is a non-Abelian two-form gauge field, which takes values in the U(4) algebra\footnote{Therefore, it can be written as $\mathcal{B}_{mn}=B_{mn}\otimes\one+\tilde{B}_{mn}^{~~~a}\otimes P_a+B_{mn}^{~~~ab}\otimes M_{ab}+B_{mn}^{~~~a}\otimes K_a+\tilde{B}_{mn}\otimes D\,.$}, as the rest of the gauge fields and transforms covariantly as:
\begin{equation}
\delta \mathcal{B}_{mn}=i[\epsilon,\hat{\Theta}_{mn}]\,,\label{covgauge}
\end{equation}
which gives $\delta\hat{\Theta}_{mn}=i[\epsilon,\hat{\Theta}_{mn}]$. From the above equation we can find the transformations of the component fields of $\mathcal{B}_{mn}$, following exactly the same procedure we describe in Appendix B for the rest of the gauge fields. Therefore, calculations lead to the following expression of the (infinitesimal) transformation of the field strength tensor:
\begin{equation}
\delta \hat{F}_{mn}=i[\epsilon,\hat{F}_{mn}]\,,\label{fieldstrengthtransformation}
\end{equation} 
that is a desired covariant transformation. 

Since, both terms in the expression of the field strength tensor transform covariantly, it is straightforward to show that the corresponding Yang-Mills 
action is gauge invariant:
\begin{equation}
\delta S = \mathrm{Tr}(\delta \hat{F}\hat{F}+ \hat{F}\delta \hat{F}) = Tr(i[\epsilon,\hat{F}]\hat{F}+i\hat{F}[\epsilon,\hat{F}])= 0\,.
\end{equation}
Therefore, gauge covariant transformation of the field strength tensor and gauge invariance of the action are ensured. 

At this point, it is also important to define the field strength tensor, $\hat{\mathcal{H}}_{mnp}$, of the 2-form gauge field:
\begin{equation}
\hat{\mathcal{H}}_{mnp}=\frac{1}{3}\left([\hat{X}_m,\hat{\Theta}_{np}]+[\hat{X}_n,\hat{\Theta}_{pm}]+[\hat{X}_p,\hat{\Theta}_{mn}]\right)\,.\label{deffieldstrengthH}
\end{equation}
In order to show that the above field strength tensor transforms covariantly under a gauge transformation, we start from the expression: 
\begin{equation}
\delta \hat{\mathcal{H}}_{mnp}=\frac{1}{3}\left([\delta\hat{X}_m,\hat{\Theta}_{np}]+[\hat{X}_m,\delta\hat{\Theta}_{np}]+[\delta\hat{X}_n,\hat{\Theta}_{pm}]+[\hat{X}_n,\delta\hat{\Theta}_{pm}]+[\delta\hat{X}_p,\hat{\Theta}_{mn}]+[\hat{X}_p,\delta\hat{\Theta}_{mn}]\right)\,
\end{equation}
and using the transformation properties of $\hat{X}_m$ and $\hat{\Theta}_{mn}$, given in equations \eqref{covcoord} and \eqref{covgauge}, respectively, and the Jacobi identity, we find:
\begin{equation}
\delta \hat{\mathcal{H}}_{mnp}=i[\epsilon,\hat{\mathcal{H}}_m]\,.
\end{equation}
Next, we expand the $\hat{\mathcal{H}}$ on the generators of the algebra:
\begin{equation}
\hat{\mathcal{H}}_{mnp}=H_{mnp}\otimes \one+\tilde{H}_{mnp}^{~~~~a}\otimes P_a+H_{mnp}^{~~~~ab}\otimes M_{ab}+H_{mnp}^{~~~~a}\otimes K_a+\tilde{H}_{mnp}\otimes D\,,\label{fieldstrengthHdecomposition}
\end{equation}
and therefore it is possible to calculate each component using the definition of the field strength tensor $\hat{\mathcal{H}}_{mnp}$:

The inclusion of this 2-form gauge field has to be prominent in the action of the theory, too. Among other terms that will be present, one has to add a kinetic term for this field:
\begin{equation}
\mathcal{S}_{\mathcal{B}}=\text{Tr}\,\text{tr}\, \hat{\mathcal{H}}_{mnp}\hat{\mathcal{H}}^{mnp}
\end{equation}
\section{Calculations of the field transformations and curvatures}

In this appendix, we present our results for the transformations of the gauge fields, the component curvatures and their transformations. In the end, we check whether our results are valid, after the consideration of the commutative limit. 

\noindent As we showed, the generators of SO(6)$\times$U(1) are encountered in a specific representation given by 4$\times$4 matrices (SO(4) notation):
\begin{equation}
\mathrm{1}, \ \ \   M_{ab}  = - \frac{i}{4} [\Gamma_{a} , \Gamma_{b} ] = - \frac{i}{2} \Gamma_{a} \Gamma_{b}, \ \ \ \frac{1}{2} \Gamma_{a}, \ \ \  -\frac{i}{2} \Gamma_{a} \Gamma_{5}, \ \ \ -\frac{1}{2} \Gamma_{5} \,.
\end{equation}
We start from an SO(5) notation and introduce the matrices $\Gamma_{A}$ with $A, B= 1\ldots 5\,,$ satisfying the well-known anticommutation relation:
\begin{equation}
\{\Gamma_{A}, \Gamma_{B} \} = 2 \delta_{AB} \one\,.
\end{equation} 
Taking this into consideration, the above generators can be written in the following more compact (SO(5) notation) form:
\begin{equation}
\mathrm{1}, \ \   \Gamma_{A}, \ \  \  M_{AB}  = - \frac{i}{4} [\Gamma_{A} , \Gamma_{B} ]\,.
\end{equation}
The generators in this SO(5) notation satisfy the following commutation and anticommutation relations \cite{Smolin:2003qu}:
\begin{eqnarray}
&&[M_{AB}, M_{CD} ] = i( \delta_{AC} M_{BD} + \delta_{BD} M_{AC} - \delta_{BC}M_{AD}  - \delta_{AD} M_{BC} )\,, \nonumber \\
&& [\Gamma_{M}, M_{NP}] = i( \delta_{MP} \Gamma_{N} - \delta_{MN}\Gamma_{P} )\,, \nonumber \\
&& \{M_{AB} , \Gamma_{C} \} = \sqrt{2}\epsilon_{ABCDE} M_{DE}\,, \nonumber \\
&& \{M_{AB}, M_{CD} \} = \frac{1}{8} ( \delta_{AC}  \delta_{BD} - \delta_{AD} \delta_{BC} )\mathrm{1} + \frac{\sqrt{2}}{8}\epsilon_{ABCDE} \Gamma_{E}\,.\label{so(5)commanticomm}
\end{eqnarray}
In turn, the covariant coordinate is written as:
\begin{equation}
\hat{X}_{m}(X) = X_{m} \otimes \one + A_{m}(X) \otimes \one + A_{m}^{~B}(X) \otimes \Gamma_{B} + A_{m}^{~AB}(X) \otimes M_{AB} \label{covariantcoordso5}
%&& \tilde{X}_{m} \otimes \mathrm{1} +  A_{m}^{B} \otimes \Gamma_{B} + A_{m}^{AB} \otimes M_{AB}\,,
\end{equation}
%with $\tilde{X}_{m} = X_{m} + A_{m} $.
and the gauge parameter as:
\begin{equation}
\epsilon(X)=\epsilon_0(X)\otimes\one+A_m(X)\otimes\one+A_m^{~A}(X)\otimes\Gamma_A+A_m^{~AB}(X)\otimes M_{AB}\,.\label{gaugeparameterso5}
\end{equation}
The definition of the field strength tensor is: 
\begin{equation} 
\hat{F}_{mn} = [\hat{X}_{m}, \hat{X}_{n}] - \frac{i\hbar}{\lambda^2}\hat{\Theta}_{mn} \otimes \one \label{fieldstrengthso5a}
\end{equation}
and is also decomposed on the generators:
\begin{equation}
\hat{F}_{mn} = F_{mn}(\one)  \otimes \one + F^{~~~A}_{mn}(\Gamma_{A}) \otimes \Gamma_{A} + F^{~~~AB}_{mn}(M_{AB}) \otimes M_{AB}\,. 
\end{equation}
After the above rewriting of the algebra and the gauge-theory related expressions, we could continue with the calculations in this SO(5) notation and then use the following decompositions in order to recover the results in the SO(4) desired language. First, the decompositions of the SO(5) generators are:
\begin{equation}
\Gamma_{A} \rightarrow \Gamma_{a} , \Gamma_{5} , \ \ \ \ M_{AB} \rightarrow M_{ab} , M_{a5}, \ \ \ B_{mn}^{~~~AB}\rightarrow B_{mn}^{~~~ab},B_{mn}^{~~~a5} , \ \ \ B_{mn}^{~~~A}\rightarrow B_{mn}^{~~~a}, B_{mn}^{~~~5}\,.
\end{equation}
and identification of the SO(5) component generators with the SO(4) ones is:
\begin{equation}
M_{ab}=-\frac{i}{4}[\Gamma_a,\Gamma_b]\,, \ \ \ M_{a5}=-\frac{i}{2}\Gamma_a\Gamma_5=P_a\,, \ \ \ \Gamma_a=2K_a\,, \ \ \ \Gamma_5=-2D\,, \ \ \ \one\,, 
\end{equation}
Accordingly, we would decompose the gauge fields to the SO(4) notation:
\begin{equation}
A_m^{~AB}\rightarrow (A_m^{~ab}\equiv\omega_m^{~ab}, A_m^{~a5}\equiv e_m^{~a})\,, \ \ \ A_m^{~A}\rightarrow (A_m^{~a}\equiv b_m^{~a}, A_m^{~5}\equiv\tilde{a}_m)\,, \ \ \ A_m\rightarrow a_m\,, 
\end{equation}
as well as the components of the SO(5) gauge parameter:
\begin{equation}
\lambda_{AB}\rightarrow (\lambda_{ab},\lambda^{a5}\equiv\tilde{\xi}^a)\,,\ \ \  \xi^A\rightarrow (\xi^a,\xi^5\equiv\tilde{\epsilon}_0) \ \ \ \epsilon_0\rightarrow \epsilon_0\,. 
\end{equation}
Instead, in order to avoid rewriting and calculating everything in the SO(5) notation, we choose the alternative (but equivalent) and more straightforward route, that is to rewrite the anticommutation relations of \eqref{so(5)commanticomm} in the SO(4) notation, taking into account the above decompositions and identifications. Then the anticommutation relations of the generators in the SO(4) notation are obtained:
\begin{align}
    \{M_{ab},M_{cd}\}&=\frac{1}{8}\left(\delta_{ac}\delta_{bd}-\delta_{bc}\delta_{ad}\right)\one-\frac{\sqrt{2}}{4}\epsilon_{abcd}D\nonumber\\
    \{M_{ab},K_c\}&=\sqrt{2}\epsilon_{abcd}P_d\,,\quad \{M_{ab},P_c\}=-\frac{\sqrt{2}}{4}\epsilon_{abcd}K_d\nonumber\\
    \{K_a,K_b\}&=\frac{1}{2}\delta_{ab}\one\,,\quad \{P_a,P_b\}=\frac{1}{8}\delta_{ab}\one\,,\quad \{K_a,D\}=\{P_a,D\}=0\nonumber\\
    \{P_a,K_b\}&=\{M_{ab},D\}=-\frac{\sqrt{2}}{2}\epsilon_{abcd}M_{cd}\,.\quad \label{anticomso(4)}
\end{align}

\noindent Making use of \eqref{covcoordinatedefinition}-\eqref{covcoordinatedecomposition} and taking into consideration the above anticommutation relations \eqref{anticomso(4)} along with the corresponding commutation relations \eqref{algebra}, the transformations of the component gauge fields are obtained:
\begin{align}
\delta\omega_m^{~ab}&=-i[X_m+a_m,\lambda^{ab}]+i[\epsilon_0,\omega_m^{~ab}]-\frac{1}{2}\{\xi^a,b_m^{~b}\}-2\{\lambda^a_{~c},\omega_m^{~bc}\}-\frac{1}{2}\{\tilde{\xi}^a,e_m^{~b}\}\nonumber\\
&\quad +\frac{\sqrt{2}i}{4}[\xi^c,e_m^{~d}]\epsilon_{abcd}-\frac{\sqrt{2}i}{4}[\tilde{\epsilon}_0,\omega_m^{~cd}]\epsilon_{abcd}-\frac{\sqrt{2}i}{4}[\lambda^{cd},\tilde{a}_m]\epsilon_{abcd}-\frac{\sqrt{2}i}{4}[\tilde{\xi}^c,b_m^{~d}]\epsilon_{abcd}\nonumber\\
\delta e_m^{~a}&=-i[X_m+a_m,\tilde{\xi}^a]+i[\epsilon_0,e_m^{~a}]+\frac{1}{2}\{\xi^a,\tilde{a}_m\}-\frac{1}{2}\{\tilde{\epsilon}_0,b_m^{~a}\}+\{\lambda^a_{~b},e_m^{~b}\}-\{\tilde{\xi}_b,\omega_m^{~ab}\}\nonumber\\
&\quad -\frac{\sqrt{2}i}{2}[\xi^b,\omega_m^{~cd}]\epsilon_{abcd}-\frac{\sqrt{2}i}{2}[\lambda^{bc},b_m^{~d}]\epsilon_{abcd}\nonumber\\
\delta b_m^{~a}&=-i[X_m+a_m,\xi^a]+i[\epsilon_0,b_m^{~a}]-\{\xi_b,\omega_m^{~ab}\}+\frac{1}{2}\{\tilde{\epsilon}_0,e_m^{~a}\}+\{\lambda^a_{~b},b_m^{~b}\}-\frac{1}{2}\{\tilde{\xi}^a,\tilde{a}_m\}\nonumber\\
&\quad +\frac{\sqrt{2}i}{8}[\lambda^{bc},e_m^{~d}]\epsilon_{abcd}+\frac{\sqrt{2}i}{8}[\tilde{\xi}^b,\omega_m^{~cd}]\epsilon_{abcd}\nonumber\\
\delta a_m&=-i[X_m+a_m,\epsilon_0]+\frac{1}{4}[\xi^a,b_m^{~a}]+\frac{i}{4}[\tilde{\epsilon}_0,\tilde{a}_m]+\frac{i}{8}[\lambda_{ab},\omega_m^{~ab}]+\frac{i}{16}[\tilde{\xi}_a,e_m^{~a}]\nonumber\\
\delta\tilde{a}_m&=-i[X_m+a_m,\tilde{\epsilon}_0]+i[\epsilon_0,\tilde{a}_m]-\frac{1}{2}\{\xi_a,e_m^{~a}\}+\frac{1}{2}\{\tilde{\xi}_a,b_m^{~a}\}-\frac{\sqrt{2}}{8}[\lambda^{ab},\omega_m^{~cd}]\epsilon_{abcd}\,.
\end{align}
The same holds for the component curvatures. The expressions of the component tensors of $\mathcal{R}_{mn}$ of  are obtained using \eqref{fieldstrengtt} and \eqref{fieldstrengthtensordecomposition}: 
\begin{align}
R_{mn}&=[X_m,a_n]-[X_n,a_m]+[a_m,a_n]+\frac{1}{4}[b_m^{~a},b_{na}]+\frac{1}{4}[\tilde{a}_m,\tilde{a}_n]+\frac{1}{8}[\omega_m^{~ab},\omega_{nab}]\nonumber\\
&~~~+\frac{1}{16}[e_{ma},e_n^{~a}]-\frac{i\lambda^2}{\hbar}B_{mn}\label{R(1)}\\
\tilde{R}_{mn}&=[X_m+a_m,\tilde{a}_n]-[X_n+a_n,\tilde{a}_m]+\frac{i}{2}\{b_{ma},e_n^{~a}\}-\frac{i}{2}\{b_{na},e_m^{~a}\}\nonumber\\
~~~&-\frac{\sqrt{2}}{8}\epsilon_{abcd}[\omega_m^{~ab},\omega_n^{~cd}]-\frac{i\lambda^2}{\hbar}\tilde{B}_{mn}\label{R(D)} \\ 
R_{mn}^{~~~a}&=[X_m+a_m,b_n^{~a}]-[X_n+a_n,b_m^{~a}]+i\{b_{mb},\omega_m^{~ab}\}-i\{b_{nb},\omega_m^{~ab}\}\nonumber \\
&\quad -\frac{i}{2}\{\tilde{a}_m,e_n^{~a}\}+\frac{i}{2}\{\tilde{a}_n,e_m^{~a}\}+\frac{\sqrt{2}}{8}\epsilon_{abcd}([e_m^{~b},\omega_n^{~cd}]-[e_n^{~b},\omega_m^{~cd}])-\frac{i\lambda^2}{\hbar}B_{mn}^{~~~a}\label{R(K)}\\
\tilde{R}_{mn}^{~~~a}&=[X_m+a_m,e_n^{~a}]-[X_n+a_n,e_m^{~a}]-\frac{i}{2}\{b_m^{~a},\tilde{a}_n\}+\frac{i}{2}\{b_n^{~a},\tilde{a}_m\}\nonumber\\
&\quad -\frac{\sqrt{2}}{2}([b_m^{~b},\omega_n^{~cd}]-[b_n^{~b},\omega_m^{~cd}])\epsilon_{abcd}-i\{\omega_m^{~ab},e_{nb}\}+i\{\omega_n^{~ab},e_{mb}\}-\frac{i\lambda^2}{\hbar}\tilde{B}_{mn}^{~~~a}\label{R(P)}\\
R_{mn}^{~~~ab}&=[X_m+a_m,\omega_n^{~ab}]-[X_n+a_n,\omega_m^{~ab}]+\frac{i}{2}\{b_m^{~a},b_n^{~b}\}+\frac{\sqrt{2}}{4}([b_m^{~c},e_n^{~d}]-[b_n^{~c},e_m^{~d}])\epsilon_{abcd}\nonumber \\
&\quad -\frac{\sqrt{2}}{4}([\tilde{a}_m,\omega_n^{~cd}]-[\tilde{a}_n,\omega_m^{~cd}])\epsilon_{abcd}+2i\{\omega_m^{~ac},\omega_{n~c}^{~b}\}+\frac{i}{2}\{e_m^{~a},e_n^{~b}\}-\frac{i\lambda^2}{\hbar}B_{mn}^{~~~ab}\label{R(M)}
\end{align}
It is very welcome that the above results of the transformations of the fields and their curvatures give the results \eqref{conformaltrans} and \eqref{conformalcurvatures}, respectively -up to some tuning of the numerical coefficients- after considering the commutative limit. In this limit, the U(1) gauge field that was introduced due to noncommutativity decouples, therefore the gauge theory in the commutative limit is the SO(6), i.e. the conformal gravity, in euclidean signature, we described in section 3. 

Eventually, we give the transformations of the component tensors, starting from eq. \eqref{fieldstrengthtransformation}, which demonstrates that the transformation of the field strength tensor of the theory is covariant. Their explicit expressions (related to the gauge invariance of the action) are:
\begin{align}
    \delta R_{mn}&=i[\epsilon_0,R_{mn}]+\frac{1}{16}[\tilde{\xi}^a,\tilde{R}_{mna}]+\frac{i}{8}[\lambda^{ab},R_{mnab}]+\frac{i}{4}[\xi^a,R_{mna}]+\frac{i}{4}[\tilde{\epsilon}_0,\tilde{R}_mn]\nonumber\\
    \delta\tilde{R}_{mn}&=i[\epsilon_0,\tilde{R}_{mn}]+\frac{1}{2}\{\tilde{\xi}^a,R_{mna}\}-\frac{\sqrt{2}i}{16}\epsilon_{abcd}[\lambda^{ab},R_{mn}^{~~~cd}]-\frac{1}{2}\{\xi^a,\tilde{R}_{mna}\}+i[\tilde{\epsilon}_0,R_{mn}]\nonumber\\
    \delta R_{mn}^{~~~a}&=i[\epsilon_0,R_{mn}^{~~~a}]+\frac{\sqrt{2}i}{8}\epsilon_{abcd}[\tilde{\xi}^b,R_{mn}^{~~~cd}]-\frac{1}{2}\{\tilde{\xi}^a,\tilde{R}_{mn}\}+\frac{\sqrt{2}i}{8}\epsilon_{abcd}[\lambda^{bc},\tilde{R}_{mn}^{~~~d}]\nonumber\\
    &~~+\{\lambda^{ab},R_{mnb}\}-\{\xi_b,R_{mn}^{~~~ab}\}+i[\xi^a,R_{mn}]+\frac{1}{2}\{\tilde{\epsilon}_0,\tilde{R}_{mn}^{~~~a}\}\nonumber\\
    \delta\tilde{R}_{mn}^{~~~a}&=i[\epsilon_0,\tilde{R}_{mn}^{~~~a}]-\{\tilde{\xi}_b,R_{mn}^{~~~ab}\}+i[\tilde{\xi}^a,R_{mn}]+\{\lambda^{ab},\tilde{R}_{mnb}\}-\frac{\sqrt{2}i}{2}\epsilon_{abcd}[\lambda^{bc},R_{mn}^{~~~d}]\nonumber\\
    &~~-\frac{\sqrt{2}i}{2}\epsilon_{abcd}[\xi^b,R_{mn}^{~~~cd}]+\frac{1}{2}\{\xi^a,\tilde{R}_{mn}\}-\frac{1}{2}\{\tilde{\epsilon}_0,R_{mn}^{~~~a}\}\nonumber\\
    \delta R_{mn}^{~~~ab}&=i[\epsilon_0,R_{mn}^{~~~ab}]-\frac{1}{2}\{\tilde{\xi}^a,\tilde{R}_{mn}^{~~~b}\}-\frac{\sqrt{2}i}{4}\epsilon_{abcd}[\tilde{\xi}^c,R_{mn}^{~~~d}]-2\{\lambda^{ac},R_{mn~c}^{~~~b}\}+i[\lambda^{ab},R_{mn}]\nonumber\\
    &~~-\frac{\sqrt{2}i}{4}\epsilon_{abcd}[\lambda^{cd},\tilde{R}_{mn}]+\frac{\sqrt{2}i}{2}\epsilon_{abcd}[\xi^a,\tilde{R}_{mn}^{~~~b}]-\frac{1}{2}\{\xi^c,R_{mn}^{~~~d}\}-\frac{\sqrt{2}i}{4}\epsilon_{abcd}[\tilde{\epsilon}_0,R_{mn}^{~~~cd}]\,.\label{transforamtionsofthecomponenttensots}
    \end{align}
\\\\ \noindent\textbf{Acknowledgements}\\
\noindent We feel that we have benefited a lot from discussions with Ali Chamseddine and Kelly Stelle. We would also like to thank Paolo Aschieri, Thanassis Chatzistavrakidis, Evgeny Ivanov, Dieter L\"ust, Denjoe O'Connor, Harold Steinacker and Christof Wetterich for useful discussions. The work of two of us (GM and GZ) was partially supported by the COST Action MP1405, while both would like to thank ESI - Vienna for the hospitality during their participation in the Workshop "Matrix Models for Noncommutative Geometry and String Theory", Jul 09 - 13, 2018. One of us (GZ) has been supported within the Excellence Initiative funded by the German and States Governments, at the Institute for Theoretical Physics, Heidelberg University and from the Excellent Grant Enigmass of LAPTh. GZ would like to thank the ITP - Heidelberg, LAPTh - Annecy and MPI - Munich for their hospitality.

\end{document}